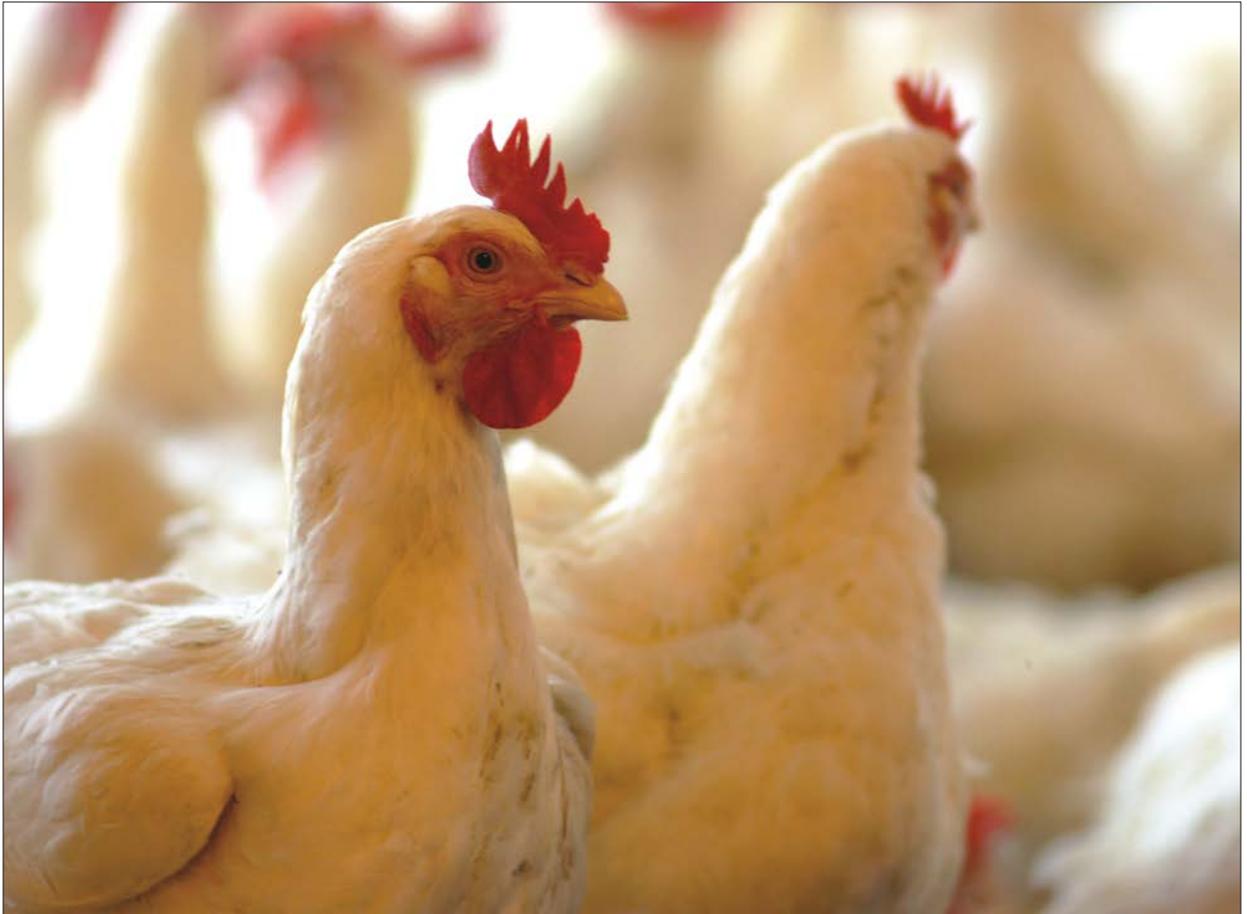

# Asian Journal of
# Poultry Science

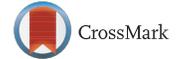

# Research Article
# Morphological and Biochemical Adaptive Changes Associated With A Short-period Starvation of Adult Male Japanese Quail (*Coturnix japonica*)

[1]Yasser A. Ahmed, [1]Soha A. Soliman and [2]Mohammed Abdelsabour-Khalaf

[1]Department of Histology, Faculty of Veterinary Medicine, South Valley University, 83523 Qena, Egypt
[2]Department of Anatomy and Embryology, Faculty of Veterinary Medicine, South Valley University, 83523 Qena, Egypt

## Abstract

**Objective:** The morphological and biochemical impact of a short-period of starvation on Japanese quail was investigated. **Materials and Methods:** Ten adult male Japanese quail were divided into two groups; control fed and starved. The control-fed group was offered food and water *ad libitum* and the starved group was subjected to a short-period of food deprivation. After 2.5 days, the serum was obtained and different parameters including the total protein, AST, ALT, triglyceride, HDL, LDL, creatinine and urea were assessed. Gastrointestinal tract, stomach and liver were excised and their masses were estimated. Paraffin and resin embedded sections from the proventriculus, gizzard, liver, duodenum, kidney and pancreas were examined with a light microscopy. **Results:** Significant decreases in the masses of body, gastrointestinal tract, stomach and liver of the starved group were recorded. The liver and duodenum were the most affected organs. The liver showed depletion of glycogen, vacuolation, hyperemia and cellular infiltrations. Duodenal villi showed degenerative changes in lamina epithelialis and cellular infiltrations in the lamina propria. Biochemical analysis revealed a decreased level of total protein, AST and ALT, increased cholesterol, triglycerides and LDL and unchanged HDL, urea and creatinine by starvation. **Conclusion:** The current study described in details the effect of short time starvation on quail organs. Time-point adaptive responses of male quail to starvation and refeeding will be investigated in future studies.

**Key words:** Starvation, quail, histology, histomorphometry, biochemistry, blood, GIT, kidney

**INTRODUCTION**

Starvation or fasting refers to a complete exclusion of all food except water[1]. Although, terminologies of starvation and fasting are interchangeably used, there is a specific definition for each of them. Starvation is the biological condition when animals are unable to eat due to an extrinsic limitation on their food supply, however fasting means that animals cannot eat due to intrinsic factors, while food is available[2]. Birds fast usually during mounting, hatching or migrating[3] or when subjected to starvation before slaughtering to improve the meat quality[4].

It was reviewed that the vertebrates are variable in bird's tolerance to starvation or fasting. For example, small animals and birds can fast for a day, whereas some snakes and frogs can remain fasting for up to 2 years[2] while, quail could starve for a 21 day period[5]. Birds are endothermic animals with a higher metabolic rate than mammals and respond quicker and stronger to starvation than do mammals. Thus, the starvation or fasting tolerance time in birds is lower than that of mammals of a similar size; for example, 31-36 h in sparrow[6] and 3-4 days in mice[7]. Birds, like mammals, progress through three metabolic phases during food starvation or fasting to provide enough energy necessary for maintaining the physiological functions of the body organs[1,2,8]. The first phase begins after the absorption of the last nutrients from the small intestine and is characterized by using carbohydrates especially liver glycogen as a main source of energy as well as mobilization of fat and rapid decrease in body mass. In the second phase, after depletion of liver glycogen, the bird oxidizes its stored lipids and the body mass slowly reduces. In the third or critical phase, the muscle protein is degraded as a fuel after depletion of most body carbohydrates and lipids and the bird loses most of its body masses and rapidly goes to death if it is not quickly refed.

Reduction of the body mass is a commonly documented response of animals and birds to starvation[9,10]. Organ decrease in size is an adaptive mechanism to save the cost of energy needed for maintaining physiological functions. However, the rate of organ mass loss differs according to energy requirements of the animal and its way of regulating the use of stored energy units during starvation[2]. Due to its expensive energy characteristics, the digestive system is the most affected organ during starvation. It was shown that masses of intestine and liver are the most dramatically reduced organs during starvation in house sparrows[10] and migrating blackcaps[9]. The decreases in the mucosal layer thickness, villus length and width, crypt depth[10,11] and morphological changes in the enterocytes[10] as well as increased apoptosis[12] are adopted histological responses during starvation of chicken. Quail has been provided as a suitable model for studying biochemical adaptation to the short[13] or long[5] food deprivation. However, no study of associated histological changes has been undertaken. The aim of the current study was to investigate the morphometric, histological and biochemical responses of domestic adult male Japanese quail after a short period of starvation (2.5 days).

**MATERIALS AND METHODS**

**Quail raising and experiment design:** Japanese quail (*Coturnix japonica*) was raised in the Quail Research Unit found by the Department of Histology, Faculty of Veterinary Medicine, South Valley University, Qena, Egypt. The experiments were approved by the Animal Experiments Committee of the Faculty of Veterinary Medicine held on January, 2015. Ten apparently healthy 50 day old male Japanese quails with a body mass of 192±1.9 g were used for the current experiment. The experiment was carried out in March, 2015. The composition of the ration provided to the birds was summarised in Table 1.

The temperature was about 25°C and the light program was about 14 h light/day. The birds were housed in steel wire cages (40×25×25 cm) with food and water provided *ad libitum* for 2 days before the experiment. On the 3rd day at 9.00 am, the bird body masses were estimated and randomly assigned to control fed (4 birds) and starved (6 birds) groups. Food and water were offered *ad libitum* to the fed control group but food was withdrawn and only water was offered *ad libitum* to the starved group for further 2.5 days (phase 1 starvation).

**Morphometric analysis:** After 2.5 days, the body mass of birds from each group was assigned and the birds were sacrificed. The gastrointestinal tract (GIT) from the stomach to the anus, stomach and liver were excised and their masses were estimated.

Table 1: Composition of the diets provided to the birds

| Ingredient | g kg$^{-1}$ |
|---|---|
| Corn | 586.6 |
| Soybean meal | 330 |
| Sunflower oil | 45 |
| lysine | 2 |
| Methionine | 2 |
| Limestone ground, 38 | 15 |
| Dicalcium phosphate | 12 |
| Salt | 4 |
| Premix | 3 |
| Antioxidant | 0.4 |

**Histological examination:** Specimens from the proventriculus, gizzard, duodenum, liver, kidney and pancreas were fixed for 3 days at 4°C in either 4% buffered paraformaldehyde for paraffin embedding or 2.5% glutaraldehyde in 4% buffered paraformaldehyde for spur's resin embedding and processed for a light microscopy as explained in details in another study[14]. Paraffin sections were stained with hematoxylin and eosin (H and E), periodic chief acid reagents (PAS) and alcian blue (AB). Resin sections were stained with 1% toluidine blue.

**Histomorphometric study:** Photomicrographs from the light microscopy were taken by a leica digital camera and some histological parameters were measured including the length of the proventricular and gizzard surface glands and the duodenal villi using image J software.

**Biochemical analysis:** Blood samples were collected from the bleeding neck after decapitation, centrifuged in a Nahita centrifuge model 2698 (GALILEO, Madrid, Spain) at 3000 rpm for 10 min to obtain a clear serum. The serum was stored at –20°C for analyses of different biochemical parameters using commercial kits and following the company provided instructions. The parameters measured were total protein, aspartate aminotransferase (AST) alanine aminotransferase (ALT), creatinine, urea, triglycerides, cholesterol and High Density Lipoprotein Cholesterol (HDLC). All these parameters were determined by DR-7000D Semi-Automatic Chemistry Analyzer (DIRUI, Changchun, China). Low Density Lipoprotein-Cholesterol (LDLC) was calculated using Friedewald' equation[15]:

$$LDLC = \text{Total cholesterol-HDL-cholesterol triglycerides}/5$$

Reagents for the total protein were from Diamond (Cairo, Egypt), ALT and AST reagents were purchased from BIOLABO SA (Maizy, France) and reagents for triglycerides, HDL cholesterol, urea and creatinine were obtained from spectrum diagnostics (Cairo, Egypt).

**Statistical analysis:** Values were expressed as Mean±SEM. The significance of difference between groups was evaluated by the one-way analysis of variance (ANOVA). The differences were considered significant when the p-value was less than 0.05.

## RESULTS

**Changes in the masses of the body and some organs:** The mass (g) of the body, GIT, stomach and liver were estimated from the control and starved groups and expressed as Mean±SE. Statistical analysis revealed that all measured parameters significantly decreased from 202.5±4.8 (body), 22.7±0.9 (GIT), 6.6±0.1 (stomach) and 5.5±0.4 (liver) in the control birds to 155±3.4 (body), 11.7±0.3 (GIT), 4.4±0.1 (stomach) and 2.7±0.1 (liver) in the starving birds. The percentages of decrease were about 23.5% (body), 48.5% (GIT), 33.3% (stomach) and 51% (liver). The mass of GIT, stomach and liver was correlated to the body mass of the bird in each group and is expressed as percentage±SE. Statistical analysis revealed that organ/body mass ratio significantly decreased from the control to starving birds as follows: 11.2±0.3 to 7.6±0.3 (GIT), 2.7±0.1 to 1.8±0.7 (liver) and 3.2±0.1 to 2.8±0.1 (stomach), respectively.

**Histological changes:** Microscopic examination revealed some histological alterations in the studied organs from the starved group comparing to the control birds. The proventriculus surface epithelium of the control group showed normal well-organized plicae and sulci lined by intact columnar epithelium (Fig. 1a), which apical surfaces stained positive for PAS (Fig. 1b). Furthermore, the proventricular glands were lined by cuboidal active oxyntic-peptic cells, separated by furrows that gave them a serrated appearance (Fig. 1c). The proventricular changes of the starved birds were as follow; extensive degenerative changes at the surface epithelium of the plicae and sulci (Fig. 1d) and loss of the PAS affinity at these areas (Fig. 1e). The oxyntic-peptic cells became more flattened and the intercellular furrows disappeared and its characteristic serrated appearance were undistinguished (Fig. 1f). The gizzard tubular glands in the control quails were lined by cuboidal epithelium (Fig. 1g) and apical surface epithelium was stained positive with PAS (Fig. 1h). The gizzard from the starved quail showed shortened tubular glands and decreased PAS affinity of the surface epithelium (Fig. 1i).

The duodenal villi of the control group were lined by a lamina epithelialis of simple columnar type (enterocytes) with goblet cells that appeared empty with H and E (Fig. 2a) and stained positive with both PAS (Fig. 2b) and alcian blue (Fig. 2c). A loose connective tissue lamina propria was underneath the epithelial layer (Fig. 2a, d) and contained lymphatic vessels "Lacteals" that extend close to the tips of the villi (2a) and duodenal glands (Fig. 2d). In the starved birds, the enterocytes and goblet cells of the villus lamina epithelialis showed extensive degenerative changes (Fig. 2e) and the goblet cells stained weekly with PAS (Fig. 2f) and AB (Fig. 2g). The lamina propria cellular infiltrations were evident in the duodenum of the starved birds (Fig. 2h).

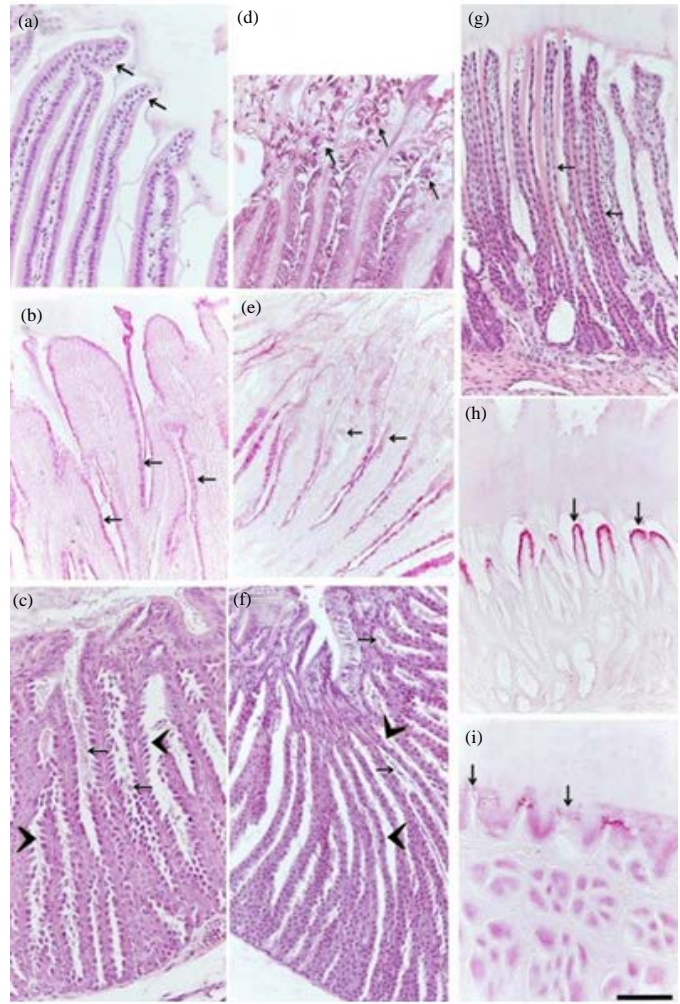

Fig.1(a-i): Effect of short-period starvation on the stomach of the quail (a-c) The proventriculus of the control quail, (d-f) The proventriculus of the starved quail, (g-h) The gizzard of the control quail and (i) The gizzard of the starved quail, (a) Normal proventricular surface epithelia of the gyri and sulci (arrows), (b) PAS-positive surface epithelium (arrows), (c) Active cuboidal oxyntic-peptic cells with a serrated appearance (arrowheads) of the proventricular glands (arrows), (d) Extensive desquamation of the surface epithelial cells (arrows), (e) Decreased PAS-affinity of the surface epithelium (arrows), (f) Flattened oxyentico-peptic cells with undistinguished serrated appearance (arrowheads) of the proventricular glands (arrows), (g) Tubular gizzard glands (arrows), (h) Gizzard PAS-positive surface epithelium (arrows), (i) Weak positive surface epithelium (arrows). (a, c, d, f, g) Paraffin sections stained with H and E and (b, e, h, i) Paraffin sections stained with PAS. All parts have the same magnification, bar = 50 μm

The liver parenchyma of the control quail consisted of plates of hepatocytes radiating from a central vein and separated by narrow blood sinusoids forming lobules, which are the structural units of hepatic tissue and no signs of inflammation could be seen (Fig. 3a-d). Glycogen contents of the hepatocytes were evidenced by intra cytoplasmic PAS staining (Fig. 3e). Most dramatic changes in the liver of starved birds were hyperemia of the vascular channels (Fig. 1g), cellular infiltration of hepatic parenchyma (Fig. 1h) and hepatocyte degeneration, vacuolation and accumulation of intracellular fat droplets (Fig. 1f, i). Furthermore, the hepatocytes of starved birds showed a week PAS stating (Fig. 1j).

The kidney of control quail consisted of cortex and medulla. The cortex contained renal corpuscles, tubules and cortical collecting ducts (Fig. 4a-d), while the medulla consisted mostly of the loop of henle (Fig. 4b). The renal corpuscles consisted of bowman's epithelial capsule and tufts

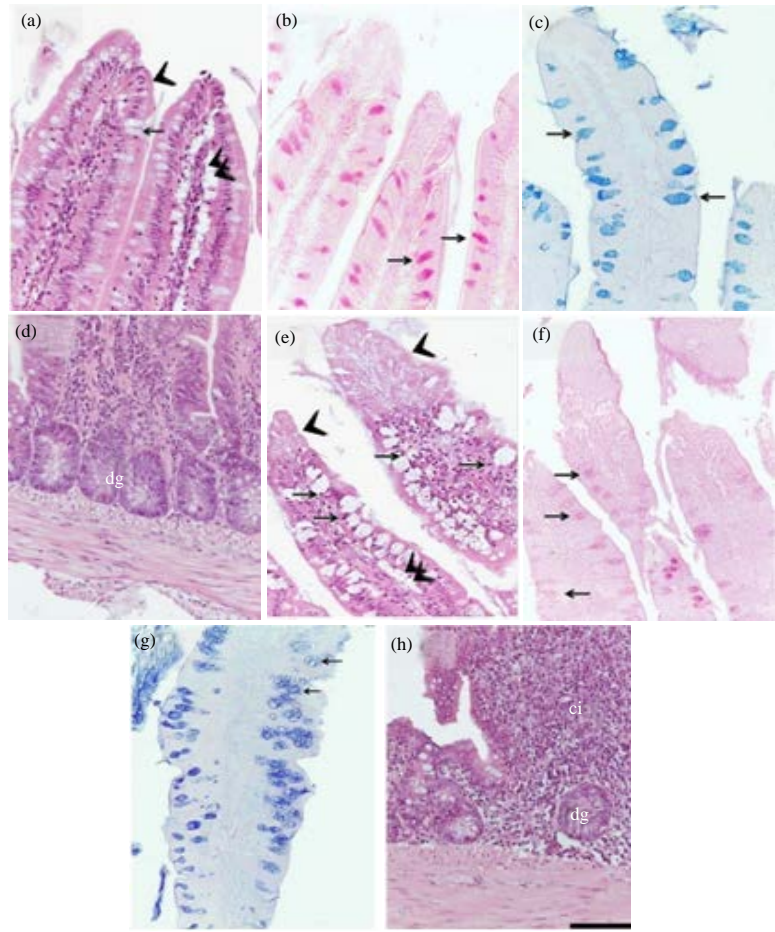

Fig. 2(a-h): Effect of short-period starvation on the duodenum of the quail (a-d) The duodenal wall of the control quail and (e-h) The duodenal wall of the starved quail, (a) Normal enterocytes (arrowheads), goblet cells (arrow) and lacteal (double arrowheads), (b) Strong PAS-positive-stained goblet cells (arrows), (c) Strong AB-positive-stained goblet cells (arrows), (d) Normal duodenal gland (dg) and lamina propria, (e) Degenerated enterocytes (arrowheads), degenerated goblet cells (arrows and shortened lacteals (double arrowheads), (f) Weak PAS-positive-stained goblet cells (arrows), (g) A decreased affinity of goblet cells to AB staining (arrows), (h) Increased cellular infiltration (ci) in the lamina propria surrounding duodenal glands (dg). (a, d, e, h) Paraffin sections stained with H and E (b, f) Paraffin sections stained with PAS and (c, g) Paraffin sections stained with AB. All parts have the same magnification, bar = 50 µm

of fine capillaries known as glomeruli (Fig. 4c). The proximal convoluted tubules were lined by cuboidal epithelium (Fig. 4d) and no evidence of cellular infiltration of the renal parenchyma was observed. In some areas maculae densa were seen as modified cells of distal consulted tubules appeared columnar in shape (Fig. 4c). In contrast, the kidney from the starved birds was identified by congestion of the capillary tufts of the glomeruli (Fig. 4e) and the blood vessels of surrounding renal parenchyma (Fig. 4f), mild degenerative changes of tubular epithelium (Fig. 4g) and cellular infiltration within the same areas of the paranchymatous tissue (Fig. 4h). The parenchyma of the quail pancreas of control birds consisted of exocrine portion (serous acini) and endocrine portion (Islets of langerhans; Fig. 4i). The pancreas of the starved birds showed mild degenerative changes in the acinar cells (Fig. 4j-l), however the endocrine portion showed no signs of morphological alterations in response to starvation.

**Histomorphometric changes:** All measured parameters from the starved group were lower than that of the control group. The length of the proventricular surface gland length, gizzard tubular glands and duodenal villi decreased from $2.5\pm0.19$, $1.6\pm0.8$ and $1.6\pm0.01$ (control fed group) to $1.5\pm0.1$, $0.31\pm0.09$ and $1.3\pm0.08$ (starved group), respectively.

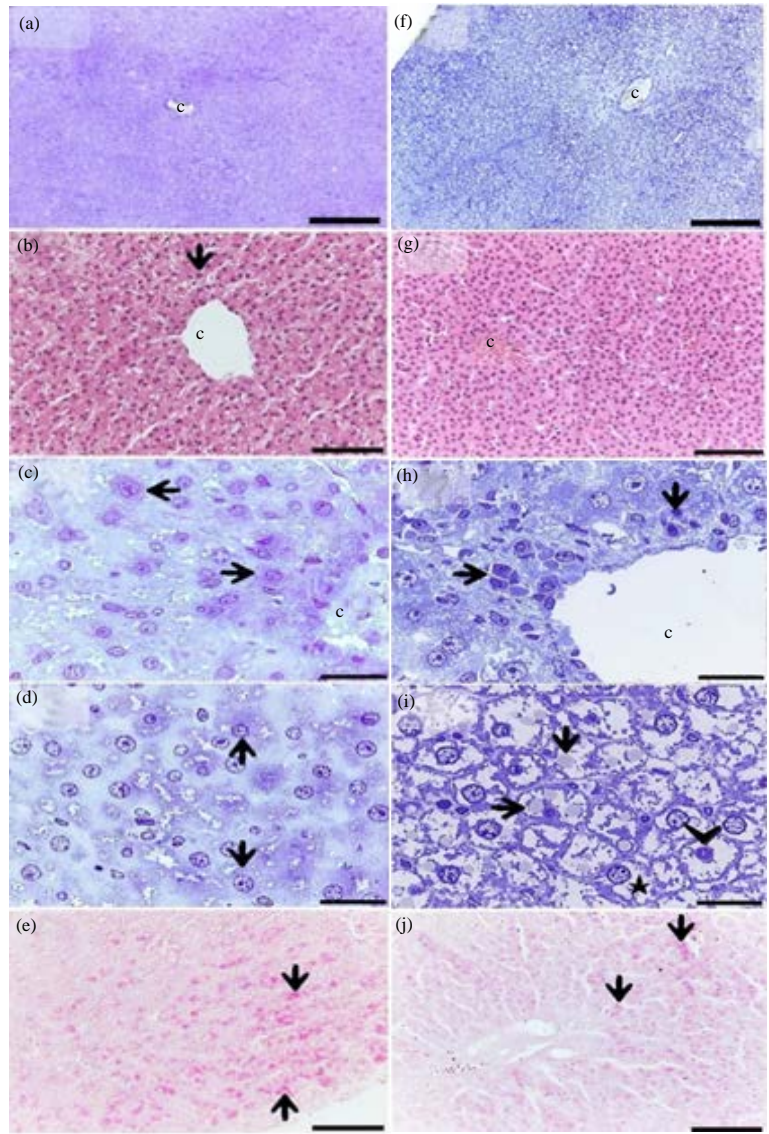

Fig. 3(a-j): Effect of short-period starvation on the liver of the quail (a-d) The liver from the control quail; note plates of hepatocytes (arrows) radiating from the central vein (c). (e) Normal glycogen contents in the hepatocytes (arrows). (f, g) Hyperemia in central veins (c), (h) Cellular infiltration (arrows) around the central vein (c). (i) Fat droplets accumulation (arrows), degeneration and vaculation (asterisk) and condensed chromatin (arrowhead) of hepatocytes, (j) Depletion of glycogen contents in the hepatocytes (arrows). (a, c, d, f, h, i) Semithin sections stained with toluidine blue and (b, e, g, j) Paraffin sections stained with H and E (b, g) Paraffin sections stained with PAS. Bars (a-f) = 8 µm, (b, g, e, i) 50 µm and (c, d, h, i) 20 µm

**Biochemical changes:** Food deprivation is associated with lowered nutrient absorption and decreased intestinal function. Serum analysis of different parameters mentioned in the materials and methods revealed that the total protein concentration significantly decreased from 5.4±0.1 in control fed group to 4.3±0.3 in the starving group. Similarly, the serum activities of AST and ALT enzymes significantly decreased from 22.5±0.7 and 45±0.9 in the control group to 17.6±0.3 and 37.5±0.6 in the fast group for AST and ALT, respectively. The calculated percentages of the decreases in total protein, AST and ALT were about 20.4, 21.7 and 16.7%, respectively. The serum level of cholesterol, triglycerides and LDL significantly increased from the control to fast groups as follows: from 112.8±1.1 to 142.5±0.8 (cholesterol), from 86±0.9 to 120±0.6 (triglycerides) and from 43.8±0.5 to 63.8±0.4 (LDL). The calculated percentages of this increase were about 26.3% (cholesterol), 39.5% (triglycerides) and 46.7% (LDL). Urea and creatinine showed no significant differences between the control and the starving group. The concentration of urea was 16.3±0.6 (control) and 16.3±0.5 (fast), while the level of creatinine was 0.53±0.03 (control) and 0.45±0.04 (fast) groups.

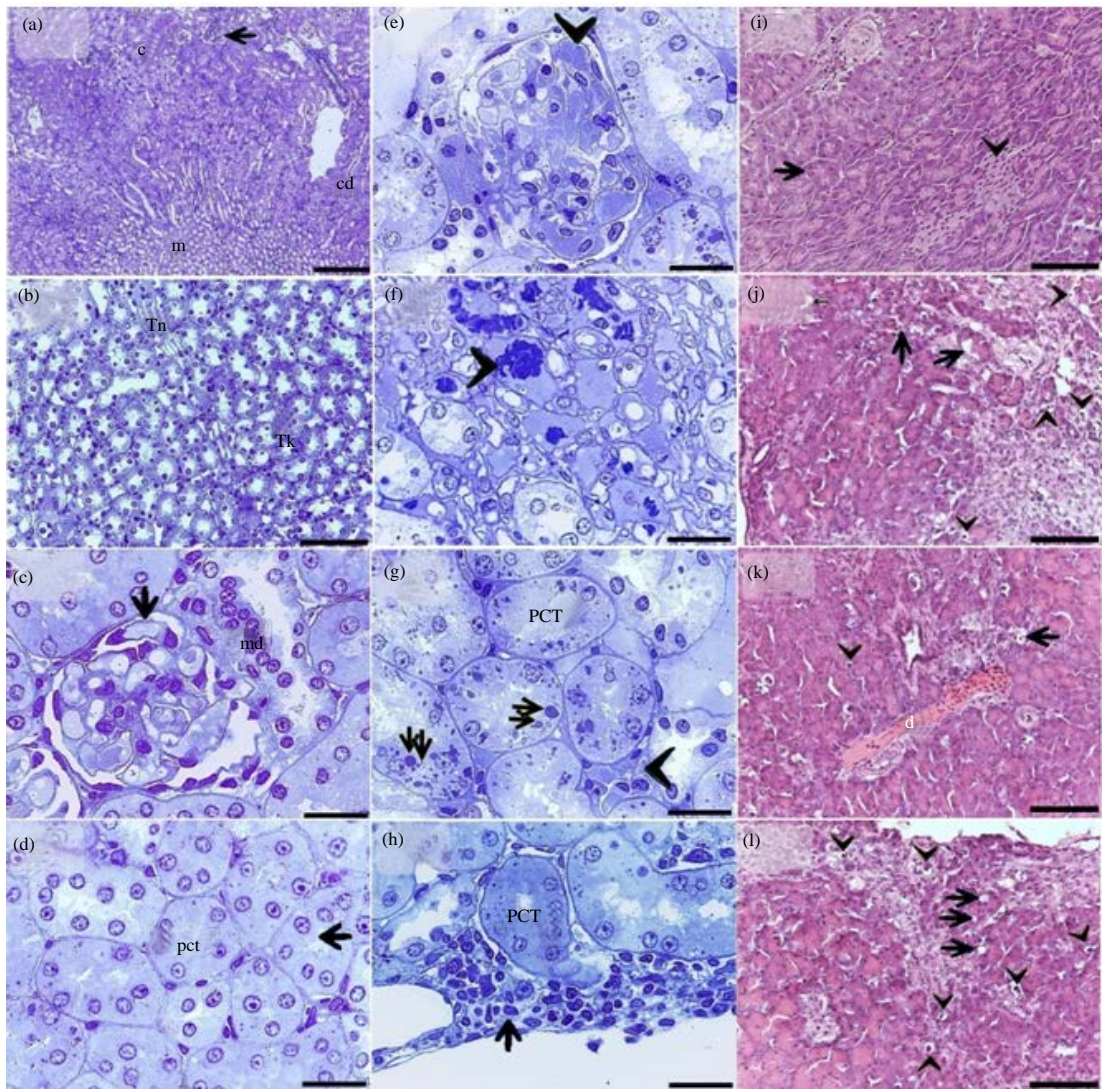

Fig. 4(a-l): Effect of short-period starvation on the kidney and pancreas of the quail (a-d) The kidney from the control quail, (a) Cortex (c), medulla (m), renal corpuscle (arrow) and cortical collecting ducts (cd), (b) Thin (Tn) and thick (Tk) segments of the loop of henle, (c) Renal corpuscle (arrow) and macula densa (md), (d) Cuboidal epithelium of the PCT (arrow). (e-h) The kidney from the starved quail, (e) Congestion of capillary tufts (arrowhead), (f) Congestion of blood capillaries of the renal parenchyma (arrowheads), (g) Degenerating epithelium (double arrows) and cellular infiltration (arrowhead) of the PCT, (h) Cellular infiltration in the renal parenchyma (arrow), (i) The pancreas from the control quail; note acini (arrow) and islets of langerhans (arrowhead) and (j-l) The pancreas from the staved quail; note mild cytoplasmic vacuolations (arrows) and nuclear condensation (arrowheads) vascular dilatation (d) in K. (a-h) Semithin sections stained with toluidine blue and parts (i-l) Paraffin sections stained with H and E. Bras (a) = 80 µm, (b, i-l) 50 µm and (c, h) 20 µm

## DISCUSSION

Similar to other avian species and mammals, quail was reported to have three phases during a 21 day long period of starvation. The first short phase (2-13) is characterized by a rabid daily loss in the body mass ranging from 6.6% at the first day then decreased to 4.2% in the subsequent 2 days. The second longest phase (about 15 days) is identified by a slow and steady loss of body mass (3-4%). The third phase (4 days) starts by the appearance of signs of weakness on the birds with rapid and higher loss of the body mass[5]. The current study was concerned with the first phase of quail starvation (2.5 days) and described the associated morphometric, histological and biochemical changes.

The current study showed that the body mass decreased in the starved birds by about 23%. Body mass loss by starvation was described in different bird species subjected to the food deprivation such as quail[5], turkey[16], penguins[17], ducks[18] and sparrows[6]. The rate of the body mass loss depends upon many factors including sex[19], avian species[6] and initial body mass[5]. The current study showed a significant decrease in measured organ masses; 48.5, 33.55 and 15% for the GAT (stomach and intestine), stomach and liver, respectively. A similar result was described in sparrows[6] and migrating blackcaps[9]. This is in agreement with the fact that the digestive system requires a high energy to perform its function[20]. The reduction of the digestive system mass could be an adaptive mechanism to decrease the high energy cost during starvation[18,21]. The decrease in the villus length described in this study is likely to be responsible for the reduction in the intestinal mass. The decrease in the villus length along with a degenerative process of the villus epithelium seen in the histology picture of the duodenum definitely reduces the absorptive power of the duodenum and thus, leading to a more decrease in the body mass in the subsequent stages. In corresponding to the current study, food deprived sparrows showed a 30% reduction in the intestinal mucosa and 20% decline in the villus length[10]. Alteration of the alimentary tract functions of quail in the current study was confirmed by the histological changes seen in proventriculus, gizzard, duodenum and liver. Furthermore, decreased PAS-stained neutral mucous and increased alcian blue-stained acidic mucus of duodenal goblet cells indicates an alteration in the mucous contents of the intestine. Increased acidic contents of goblet cells in the duodenal wall could be one of the causes of intestinal epithelium degeneration. The increased cellular infiltration seen in the duodenal mucosa in this study is an immunological response to increased intestinal cell degeneration[22].

The second most dramatic changes were seen in the liver, where hepatocytes showed degenerative changes and a lower affinity to PAS indicating a decreased glycogen content of the liver. The alteration in glycogen contents has been described in the liver of starved birds[21,23]. Depletion of the liver glycogen; glycolysis[5]. It is known that the food deprived bird is able to maintain unchanged blood glucose level during[5,23] by either glycogenolysis or gluconeogenesis. Due to alteration in the absorption of amino acids from the intestine, the gluconeogenesis is likely to decrease and glycogenolysis is activated. It was shown that pancreas showed only a few histological alterations in exocrine portion but its endocrine portion was histologically normal and likely to play its role in the regulation of insulin/glucagon pathway to help in maintaining quail blood glucose during starvation. Kidney histology showed minimal histological alteration and that was also supported by the normal level or uric acid and creatinine showed in the biochemical analysis.

The biochemical analysis of the serum revealed a decreased level of total protein by starvation. This result is in agreement of a 6 day starvation in ducklings[24]. Protein decrease is likely due to a decrease in the biosynthesis or an increase in the breakdown. If protein decrease was due to its breakdown, the uric acid and creatinine, indicators for protein destruction[25] would be higher in the starved group, however, that was not the case. Thus, the decreased total protein by starvation is likely due to a decreased biosynthesis. The shortage of amino acid supplied from starvation and absorption from the altered intestinal villi results in a lower level of circulating amino acid. Furthermore, circulating amino acid will be saved and not used for more protein synthesis but may enter a gluconeogenesis cycle to maintain normal blood glucose level. The AST and ALT decreased in starved quail. This may be due to a decreased protein regeneration caused by the lack of amino acid absorption from the intestine[24]. The plasma level of lipids (LDL, cholesterol and triacylglycerol) increased in the short-period starved quail as shown in this study. It was reported that during starvation, mobilization of lipids from its store increased to be used as a source of energy after depletion of glycogen[26]. The HDL remained unchanged after short-period starvation of quail. The HDL is the good cholesterol that never deposits on the cell membrane.

## CONCLUSION

In the current study, the morphological and biochemical changes in adult male quail resulting from 2.5 days of starvation were described. Studies should be continued to investigate changes after different time points of starvation and re-feeding.

**ACKNOWLEDGMENTS**